\documentclass{aastex7}
\usepackage{amsmath}
\shorttitle{simesh: Native AMR Analysis}
\shortauthors{Wu et al.}
\hypersetup{pdftitle={simesh: A Python Framework for Quantitative Analysis of Native AMR Data from MPI-AMRVAC},
pdfauthor={Hao Wu, Jinhan Guo, Rony Keppens, Yang Guo}}
\begin{document}
\title{simesh: A Python Framework for Quantitative Analysis\\of Native AMR Data from MPI-AMRVAC}
\correspondingauthor{Hao Wu}
\author[0009-0006-9249-8468,gname=Hao,sname=Wu]{Hao Wu}
\affiliation{State Key Laboratory of Solar Activity and Space Weather, School of Aerospace, Harbin Institute of Technology, Shenzhen 518055, China}
\email[show]{wuhao2026@hit.edu.cn}
\author[0000-0002-4205-5566,gname=Jinhan,sname=Guo]{Jinhan Guo}
\affiliation{State Key Laboratory of Solar Activity and Space Weather, School of Aerospace, Harbin Institute of Technology, Shenzhen 518055, China}
\email{jinhan.guo@kuleuven.be}
\author[0000-0003-3544-2733,gname=Rony,sname=Keppens]{Rony Keppens}
\affiliation{Centre for Mathematical Plasma Astrophysics, Department of Mathematics, KU Leuven, Celestijnenlaan 200B, 3001 Leuven, Belgium}
\email{rony.keppens@kuleuven.be}
\author[0000-0002-9293-8439,gname=Yang,sname=Guo]{Yang Guo}
\affiliation{School of Astronomy and Space Science and Key Laboratory of Modern Astronomy and Astrophysics, Nanjing University, Nanjing 210023, China}
\email{guoyang@nju.edu.cn}
\begin{abstract}
Adaptive mesh refinement (AMR) is widely used in large-scale astrophysical simulations, but the resulting multilevel data are challenging to analyze quantitatively. We present \texttt{simesh}, an independent Python framework for quantitative analysis of cell-centered fields on block-structured AMR meshes. The current implementation targets three-dimensional Cartesian datasets from MPI-AMRVAC and compatible AGILE snapshots, preserving the native AMR hierarchy and connecting complementary scientific diagnostics within a common numerical representation. The framework combines Python workflow interfaces with compiled Cython kernels, supporting efficient analysis on AMR meshes. We compare native AMR analysis with a VTK point-data conversion and uniform-grid resampling, which shows that changes in data association, interpolation support, and differentiation scale affect both sampled magnetic fields and derived current densities. Small differences in the sampled primary field can become more pronounced in derivative-based quantities, particularly near coarse--fine interfaces. The applications combine field diagnostics, tracing, regional statistics, and line-of-sight integration on the same native AMR representation. \texttt{simesh} makes these operations available through reusable Python interfaces for reproducible analysis of multiscale simulation data.
\end{abstract}

\section{Introduction}
\label{sec:introduction}
Numerical simulations provide access to the evolving physical fields of astrophysical systems, enabling the study of processes that are difficult to constrain through observations alone. Hydrodynamic and magnetohydrodynamic (MHD) simulations are widely used to study accretion, outflows, and magnetized plasma dynamics, where localized structures coexist with large-scale flows. Adaptive mesh refinement (AMR) addresses this separation of scales by increasing spatial resolution locally where required \citep{BergerOliger1984,BergerColella1989}. Compared with a uniform grid at the same finest resolution, AMR can reduce computational and memory costs while retaining fine structures within a larger domain.

Astrophysical simulation codes differ both in their mesh organization and in how their AMR infrastructure is provided. RAMSES \citep{Teyssier2002} uses a tree-based, cell-by-cell refinement strategy, whereas BATS-R-US \citep{Powell1999}, NIRVANA \citep{Ziegler2008}, Athena++ \citep{Stone2020}, and Message Passing Interface Adaptive Mesh Refinement Versatile Advection Code (MPI-AMRVAC) \citep{Porth2014,Xia2018,Keppens2023} use block-structured AMR. External libraries can also supply AMR support. FLASH \citep{Fryxell2000} employs the PARAMESH toolkit \citep{MacNeice2000}, and the AMR extension of PLUTO \citep{Mignone2012} couples its hydrodynamic and MHD solvers to the Chombo infrastructure. These implementations provide different routes to adaptive resolution, with corresponding differences in mesh hierarchies and data organization.

Current developments also extend AMR simulations to graphics processing units (GPUs). Examples include AthenaK \citep{Stone2026AthenaK}, which builds on Athena++ to provide portable execution on CPUs and GPUs, and AGILE \citep{Porth2026AGILE}, a GPU-oriented code that inherits the block-based AMR mesh handling of MPI-AMRVAC. As simulation capabilities develop, extracting physical insight from their multilevel outputs remains a distinct numerical task. Quantitative post-processing must identify resolved structures, measure their properties, and connect them to observable signatures while preserving the locally resolved scales. This requires numerical and data-handling methods that remain consistent with the AMR hierarchy.

General-purpose tools such as ParaView \citep{Ahrens2005}, VisIt \citep{Childs2012}, and \texttt{yt} \citep{Turk2011} provide scalable visualization and quantitative analysis across different simulation codes. The numerical representation used by an analysis depends on the selected reader and processing operations. Some routes retain the AMR hierarchy, whereas others convert cell-centered fields to mesh vertices or resample them onto uniform grids. Quantitative diagnostics on AMR data can therefore depend on both the chosen representation and how neighboring values are supplied to numerical stencils.

Specialized solar-analysis packages address complementary scientific tasks. FastQSL~2 \citep{Chen2026} combines QSL, magnetic-twist, and field-line diagnostics on structured Cartesian and spherical grids. Radiation Synthesis Tools (RST) \citep{Fu2026RST} provides multiwavelength emission and absorption synthesis from arbitrary viewing perspectives on Cartesian and spherical grids. These packages organize their capabilities around magnetic connectivity or radiation synthesis. Relating magnetic structures to their thermal content and observable emission also requires carrying selected regions, tracing seeds, and sampling paths between these analyses. Combining these tasks on the native block-structured AMR mesh requires a shared refinement hierarchy and local numerical support for interpolation and derivatives. This motivates a shared native-AMR representation connecting field diagnostics, tracing, regional statistics, and synthetic observations.

Here we focus on three-dimensional Cartesian, block-structured MPI-AMRVAC datasets and compatible AGILE outputs. We present \texttt{simesh}\footnote{Source code, documentation, and examples are available at \url{https://github.com/Astery0502/simesh}.}, an independent Python framework that brings quantitative diagnostics and scientific-analysis workflows onto a common native AMR basis. The aim is to relate resolved structures, their physical properties, and synthetic observables while retaining the simulation's local resolution. In the following sections, we describe the computational framework, examine its numerical behavior, and demonstrate representative applications.

\section{Software framework and numerical methods}
\label{sec:framework}

\subsection{Supported data and scientific workflows}
\label{sec:functional-organization}

\texttt{simesh} operates on three-dimensional Cartesian block-structured AMR data from MPI-AMRVAC and compatible AGILE outputs. It organizes cell-centered fields on the original block hierarchy, retaining refinement levels and local cell scales. General operations select field components by name, independently of the simulation physics module, with scalar, vector, and tensor quantities represented by their named components. Physical assumptions and units are specified explicitly when constructing model-dependent diagnostics. The current reader exposes cell-centered variables in the MPI-AMRVAC v5 format, and the array interface accepts named components on a compatible native mesh.

The framework rebuilds mesh geometry and block connectivity from snapshot metadata and prepares ghost-cell values for interpolation and differential stencils. Reconstruction schemes and differential stencils are specified explicitly, allowing comparisons with solver-side diagnostics under matched numerical settings. Analyses operate on resolved AMR cells, with resampling where required by a diagnostic or output format.

Four groups of scientific applications share these fields and the native AMR representation, comprising field diagnostics, statistics and regional integrals, tracing and topology analysis, and synthetic observations. Their outputs can be combined in workflows from structure identification and physical characterization to synthetic observables.

Field diagnostics combine pointwise expressions with spatial derivatives of selected continuous components. For example, the velocity-gradient tensor $A_{ij}=\partial v_i/\partial x_j$ can be used to construct strain-rate and rotation tensors, while $\boldsymbol{\omega}=\nabla\times\mathbf{v}$ gives the vorticity. These quantities are constructed from component-wise derivatives using the local mesh spacing and valid neighboring values. Their components and derived scalars can then be sampled or passed to statistics and regional integrals.

Within these three-dimensional datasets, users can extract native axis-aligned slices, sample arbitrarily oriented cutting planes and prescribed curves, and select subvolumes for statistics and volume integrals. Fields can also be interpolated at supplied points on a surface, including an isosurface. The sampling geometry specifies where quantities are evaluated, while the original mesh supplies their local resolution and neighboring values.

Region selection can also support a subsequent simulation. For this purpose, \texttt{simesh} reorganizes complete root-block subtrees into an independent MPI-AMRVAC \texttt{.dat} snapshot while preserving refinement levels and cell-centered field values. The extracted snapshot can initialize a localized restart simulation, whose subsequent snapshots can support test-particle studies. This approach has previously been explored by \citet{Wu2025}.

Streamlines and field lines are obtained by tracing curves tangent to a supplied three-component vector field, which may be the velocity, magnetic field, current density, or electric field. Velocity streamlines can be combined with the local derivative diagnostics to study flow topology and vortical structures. Along these curves, \texttt{simesh} can sample other fields, integrate scalar quantities during tracing, and compute line integrals, weighted averages, and derivatives with respect to arc length from sampled profiles.

\begin{figure*}[!tp]
\centering
\includegraphics[width=0.85\textwidth]{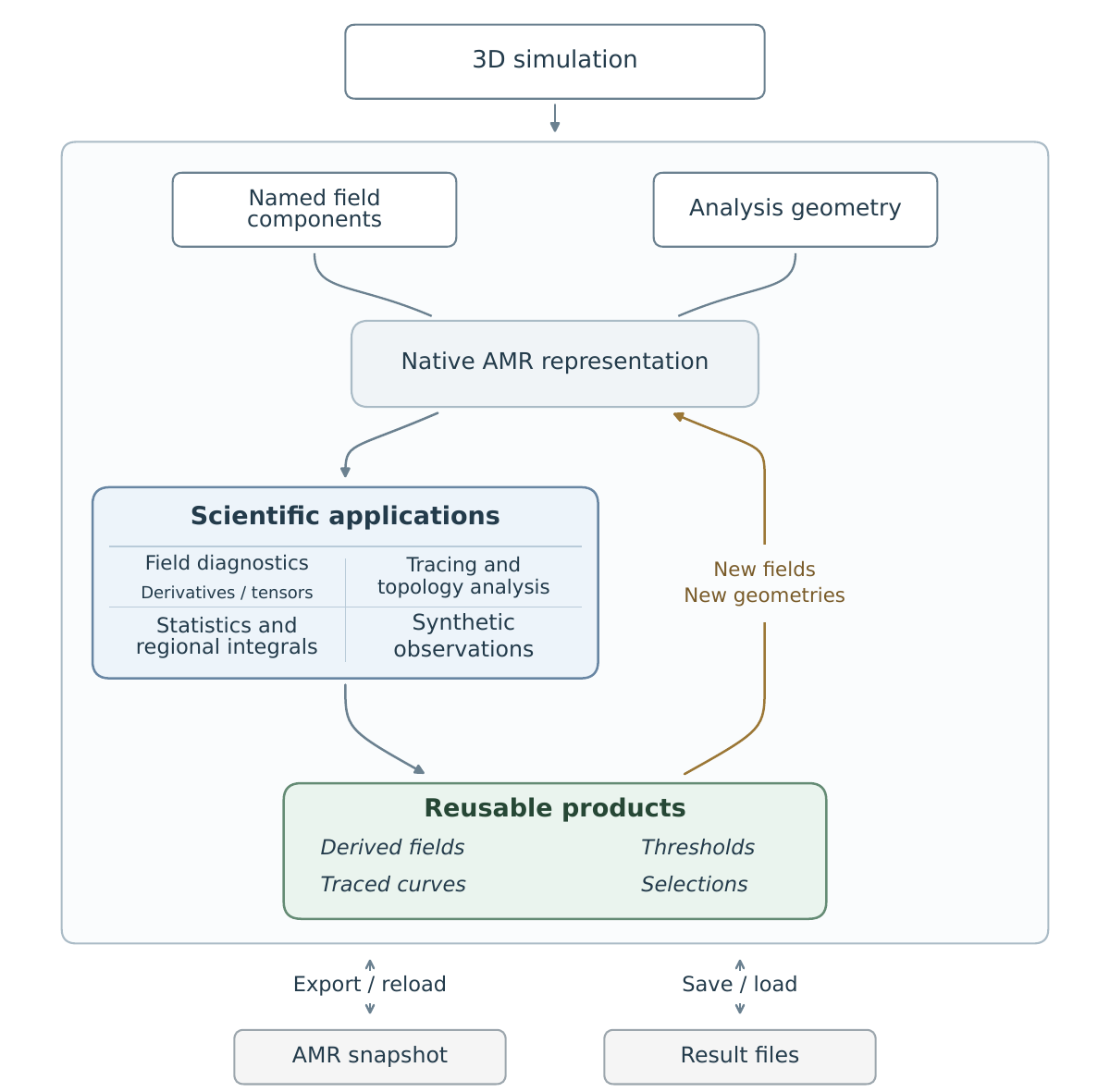}
\caption{
Overview of the \texttt{simesh} analysis framework for three-dimensional Cartesian AMR data. Named field components and analysis geometries feed four application groups through the native AMR representation, independently of the simulation physics module. Field diagnostics include spatial derivatives and tensor quantities, while tracing supports flow and magnetic-topology analysis. The resulting fields and geometries can be reused in subsequent analysis steps.
}
\label{fig:framework}
\end{figure*}

Fig.~\ref{fig:framework} summarizes these connections between simulation fields, sampling geometries, and scientific applications. Results can be reused as new fields or geometries. For example, a thresholded diagnostic can select new tracing seeds, and a traced curve can define an along-line sampling path. These connections support repeated analysis on the same resolved AMR data.

\subsection{Native AMR representation and interfaces}
\label{sec:native-interfaces}

\begin{figure*}[!tp]
\centering
\includegraphics[width=\textwidth]{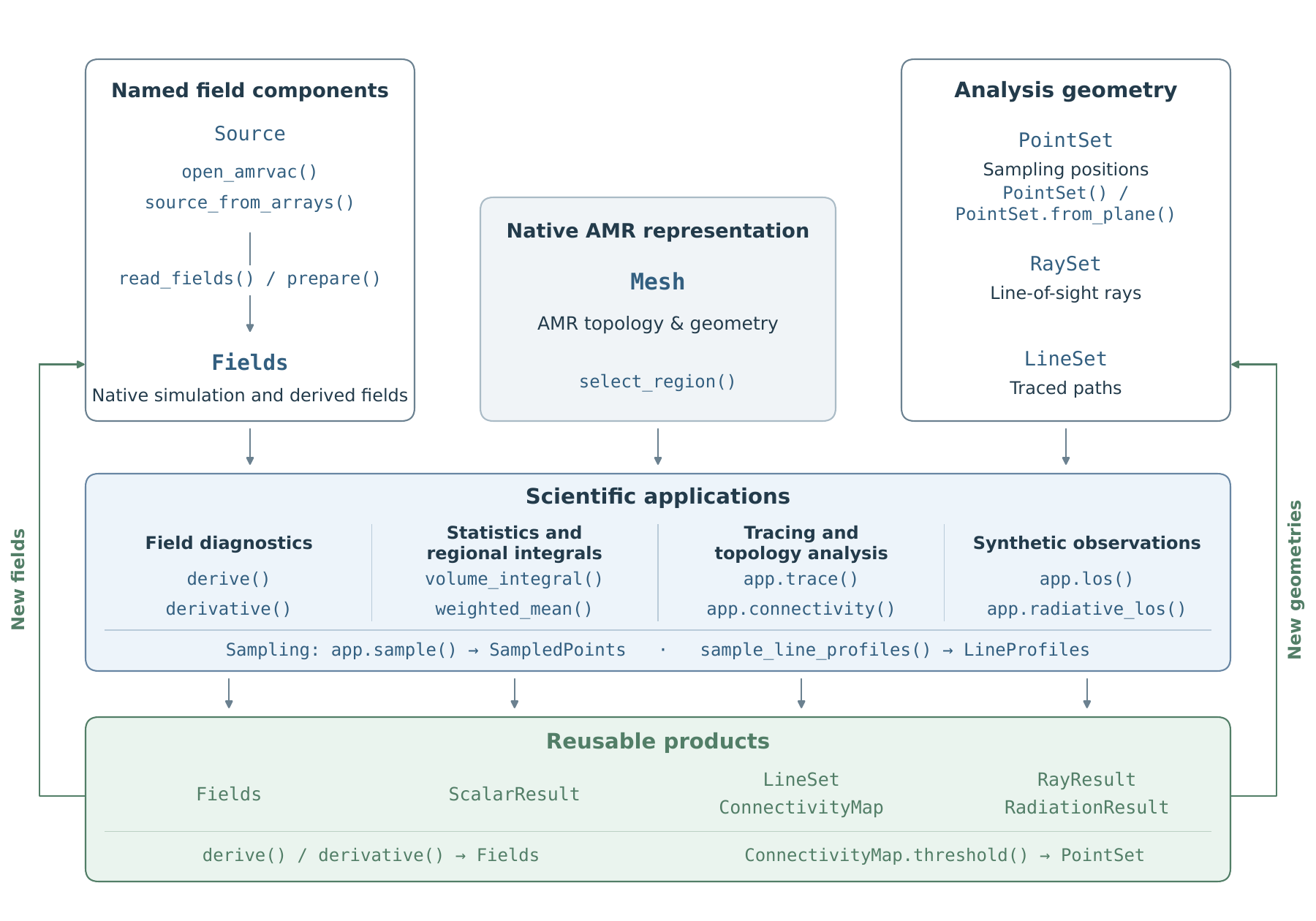}
\caption{
Native AMR representation and representative interfaces in \texttt{simesh}. \texttt{Source} supplies simulation fields, while independent geometry objects specify sampling positions, rays, or traced paths. Scientific applications combine these inputs through the native \texttt{Mesh}; derived fields and geometries can be reused in later calculations.
}
\label{fig:native-interfaces}
\end{figure*}

Fig.~\ref{fig:native-interfaces} shows how the framework is implemented through the main \texttt{simesh} interfaces, linking field data, analysis geometry, the native \texttt{Mesh}, and scientific applications.  Simulation data enter through a common \texttt{Source} interface, with \texttt{open\_amrvac()} and \texttt{source\_from\_arrays()} providing file and array inputs, respectively. For a selected mesh region, \texttt{read\_fields()} reads the requested components into \texttt{Fields}, while \texttt{prepare()} additionally constructs the valid neighboring values required by numerical stencils.  

Analysis geometry is represented independently of the field values. A \texttt{PointSet} represents physical sampling positions, including points on curves or surfaces, and may retain a regular image layout. A \texttt{Plane} defines an arbitrarily oriented sampling plane within the three-dimensional domain. A \texttt{RaySet} represents origins, directions, and integration limits for line-of-sight (LOS) calculations, and a \texttt{LineSet} stores curves traced through a chosen vector field together with their termination states. The same geometry can therefore be applied to different field quantities without redefining the underlying mesh representation.

The native \texttt{Mesh} stores a balanced octree forest in which each refined parent has eight children, cell spacing decreases by a factor of two per level, and touching leaf blocks differ by at most one level. Leaves contain regular cell arrays and cover the domain without overlap. Morton (Z-order) is the sole space-filling curve used in the current implementation, following MPI-AMRVAC \citep{Keppens2012}. 

Analysis bounds and storage coverage are distinct. \texttt{select\_region()} accepts an arbitrary axis-aligned Cartesian box, whose faces need not coincide with block or cell boundaries, and selects every intersecting complete leaf. Regional statistics and integrals use the requested box, weighting intersected cells by their overlap volumes under a piecewise-constant cell-value representation. Preparing this selection obtains neighboring values from the original mesh; a selection edge is not treated as a physical boundary.

Independent snapshot export has a stricter geometric requirement. \texttt{crop\_amrvac()} and \texttt{write\_amrvac()} export rectangular collections of complete root-block subtrees as native \texttt{.dat} files. They rebuild local tree indices and domain metadata while preserving cell locations, refinement levels, and field values, without resampling.

Fields and analysis geometry are combined through this common native AMR representation. Physical query positions are associated with their containing leaf blocks and corresponding local coordinates, allowing sampling and other mesh-dependent calculations to access the field values at the resolution of the local mesh. Interpolation and differential operators use the explicitly prepared neighboring values required by their stencils, within the valid stencil support of each field. Valid ghost-cell layers contain the neighboring values required by the selected numerical scheme. Their number describes the available stencil support. Operations such as differentiation may reduce this support for subsequent calculations.

The scientific applications are built on this shared representation. The four groups of scientific applications combine a small set of common numerical tasks, including field evaluation and differentiation, spatial sampling, regional reduction, line integration and differentiation, and line-of-sight integration. Fig.~\ref{fig:native-interfaces} lists representative user-facing interfaces for these tasks.

\subsection{Ghost-cell reconstruction and numerical support}
\label{sec:numerical-operations}
\label{sec:native-amr-structure}

\begin{figure*}[!tp]
\centering
\includegraphics[width=\textwidth,height=0.68\textheight,keepaspectratio]{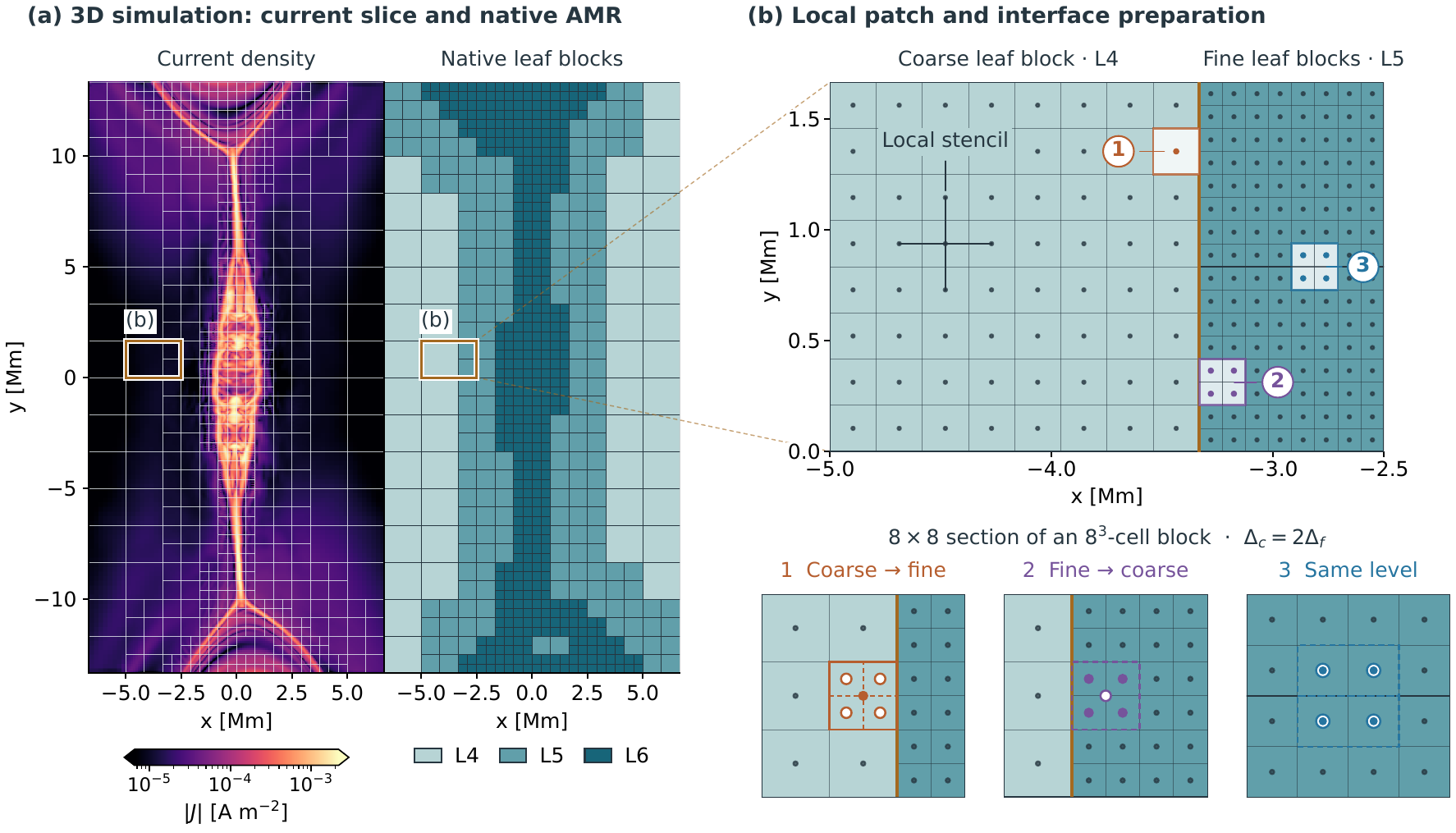}
\caption{
Native AMR structure and ghost-cell reconstruction in \texttt{simesh}.
(a) A slice at $z=5.052\,\mathrm{Mm}$ selected to visualize the AMR structure of the three-dimensional simulation of \citet{Wu2025}, showing the current-density magnitude and corresponding leaf-block layout.
(b) A two-dimensional section through $8^3$-cell blocks. The three lower enlargements show coarse-to-fine, fine-to-coarse, and same-level transfers; filled markers denote native donor values and open markers denote receiving ghost-cell centers. Out-of-plane stencil values are omitted from this schematic.
}
\label{fig:native-amr-structure}
\end{figure*}

Fig.~\ref{fig:native-amr-structure} illustrates how the native AMR representation in \texttt{simesh} connects the simulated structure to subsequent numerical analysis. To make the AMR structure of the three-dimensional current-sheet simulation studied by \citet{Wu2025} easier to visualize, Fig.~\ref{fig:native-amr-structure}(a) shows the current-density magnitude and corresponding MPI-AMRVAC leaf-block layout on a slice at $z=5.052\,\mathrm{Mm}$. The refinement is strongly nonuniform, with fine blocks concentrated around the structured current layer and coarser blocks covering the surrounding domain. This adaptive distribution preserves high spatial resolution where needed without representing the full computational domain at the finest level.

Fig.~\ref{fig:native-amr-structure}(b) shows an in-plane view of the three-dimensional ghost-cell construction. Same-level neighbors supply direct copies. Restriction averages $2\times2\times2$ fine cells, and coarse-to-fine transfer uses linear reconstruction with minmod-limited directional slopes, following MPI-AMRVAC's default cell-centered prolongation \citep{Keppens2023}.

Reconstructing ghost cells during post-processing avoids storing these layers in each snapshot, particularly reducing file volume for small blocks. Saving $g$ layers on each side of a block containing $n^3$ interior cells multiplies the uncompressed field-data volume by $(n+2g)^3/n^3$. Two layers per side therefore add about 95\% for $16^3$ blocks and 238\% for $8^3$ blocks. In \texttt{simesh}, ghost cells are prepared on demand for selected fields and regions, with an explicitly chosen reconstruction scheme. The reconstructed values still contribute to the field storage held in memory during analysis.

\subsection{Execution time and memory use}
\label{sec:execution-performance}

To quantify the computational overhead of hierarchical field access, we compare magnetic-field-line tracing on native AMR with two global uniform grids, Uniform L5 and Uniform L6, whose cell spacings correspond to AMR refinement levels 5 and 6, respectively. Field-line tracing provides a test of spatially irregular field access across mesh blocks. All configurations use the same RK4 integrator, thread-pool implementation, dynamic scheduling, and final-state output handling.

The magnetic-field snapshot shown in Fig.~\ref{fig:native-amr-structure}(a) comprises 22,614 leaf blocks of $8^3$ cells, totaling 11.58 million native cells. The Uniform L5 and L6 grids contain $256\times256\times128$ and $512\times512\times256$ cells, respectively. Both are constructed by trilinearly interpolating the native cell-centered magnetic field onto the uniform grids and are stored as contiguous $(n_x,n_y,n_z,3)$ arrays. Uniform-grid sampling uses direct index computation and trilinear interpolation, with constant continuation at physical boundaries. Native AMR sampling uses two prepared ghost-cell layers and reuses the containing leaf block when successive evaluations remain within its bounds.

Benchmarks were conducted on 1--32 physical cores of an Intel Xeon Gold 6448Y processor. Each configuration traced 65,536 seeds in both directions with a fixed step size of $1/1536$ in code units, retaining only final states. Magnetic-field values were stored in double precision. Each configuration ran in a separate process, with one warm-up followed by three timed calls. We report median wall times and maximum process RSS sampled at 50\,ms intervals. Timings include integration and output allocation but exclude input reading, field preparation, and external result checks.

\begin{figure*}[t]
\centering
\includegraphics[width=\textwidth]{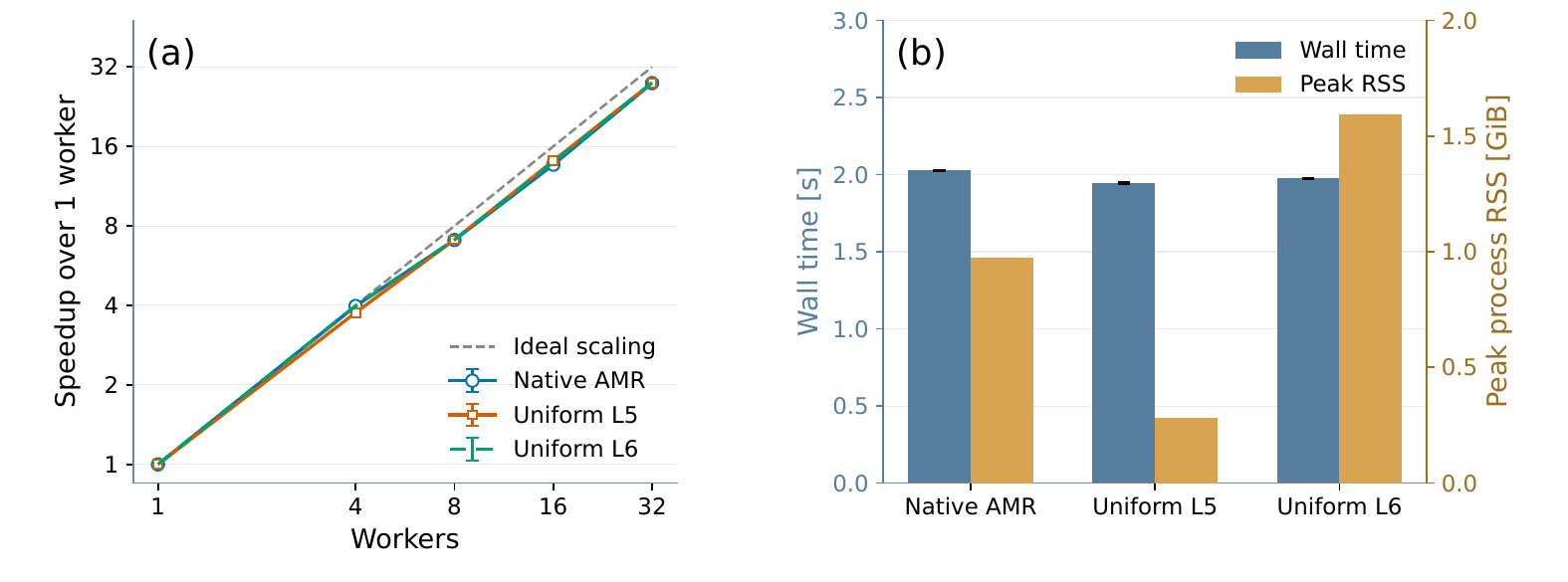}
\caption{Magnetic-field-line tracing performance for native AMR, Uniform L5, and Uniform L6. (a) Strong scaling relative to the one-worker median of each representation; the dashed line indicates ideal scaling. (b) Median wall time (blue, left axis) and peak process RSS (gold, right axis) at 32 workers, plotted on independent axes. }
\label{fig:enine-tracing-performance}
\end{figure*}

As shown in Fig.~\ref{fig:enine-tracing-performance}(a), all three representations exhibit nearly identical strong scaling. At 32 workers, native AMR, Uniform L5, and Uniform L6 achieve speedups of 27.77, 27.85, and 27.87, respectively, with corresponding median wall times of 2.03, 1.94, and 1.98\,s [Fig.~\ref{fig:enine-tracing-performance}(b)]. Native AMR therefore incurs only a small runtime overhead relative to direct uniform-grid access. Its peak RSS is 0.976\,GiB, compared with 0.282\,GiB for Uniform L5 and 1.595\,GiB for Uniform L6. Although Uniform L5 requires less memory, it does not preserve the finest native cell spacing. Native AMR thus retains locally refined resolution with substantially lower memory consumption than Uniform L6 and comparable tracing performance.

\subsection{Effects of data representation}
\label{sec:representation-diagnostics}
\subsubsection{Comparison setup}

To assess how the choice of data representation affects quantitative analysis, we compare several processing routes applied to the same magnetic-field snapshot. The four routes are denoted Native AMR, Visualization Toolkit (VTK) PointData, Uniform L5, and Uniform L6. The latter two use \texttt{simesh} to resample the cell-centered magnetic field onto uniform target grids with spacings corresponding to refinement levels L5 and L6, respectively.  The routes differ in how the field components are represented and processed before the final diagnostic is evaluated. We examine their effects on both the sampled magnetic-field component $B_x$ and a derivative-based quantity, the magnitude of the three-dimensional current density,
\begin{equation}
\mathbf{J}
=
\frac{1}{\mu_0}\nabla\times\mathbf{B},
\qquad
J=|\mathbf{J}|.
\label{eq:current-definition}
\end{equation}
Here, \(\mu_0\) is the vacuum permeability, $\mathbf{J}$ is the current-density vector, and $J$ denotes its scalar magnitude. We use the Native AMR route as the comparison baseline to examine differences introduced by data conversion, resampling, differentiation, and sampling.

A \texttt{yt} pathway was also examined. In the tested \texttt{yt} 4.4.2 MPI-AMRVAC frontend, smoothed ghost-cell reconstruction falls back to a piecewise-constant covering grid. The quantitative comparison below therefore considers the Native AMR, VTK PointData, Uniform L5, and Uniform L6 routes.

\begin{figure*}[t]
    \centering
    \includegraphics[width=\textwidth]
    {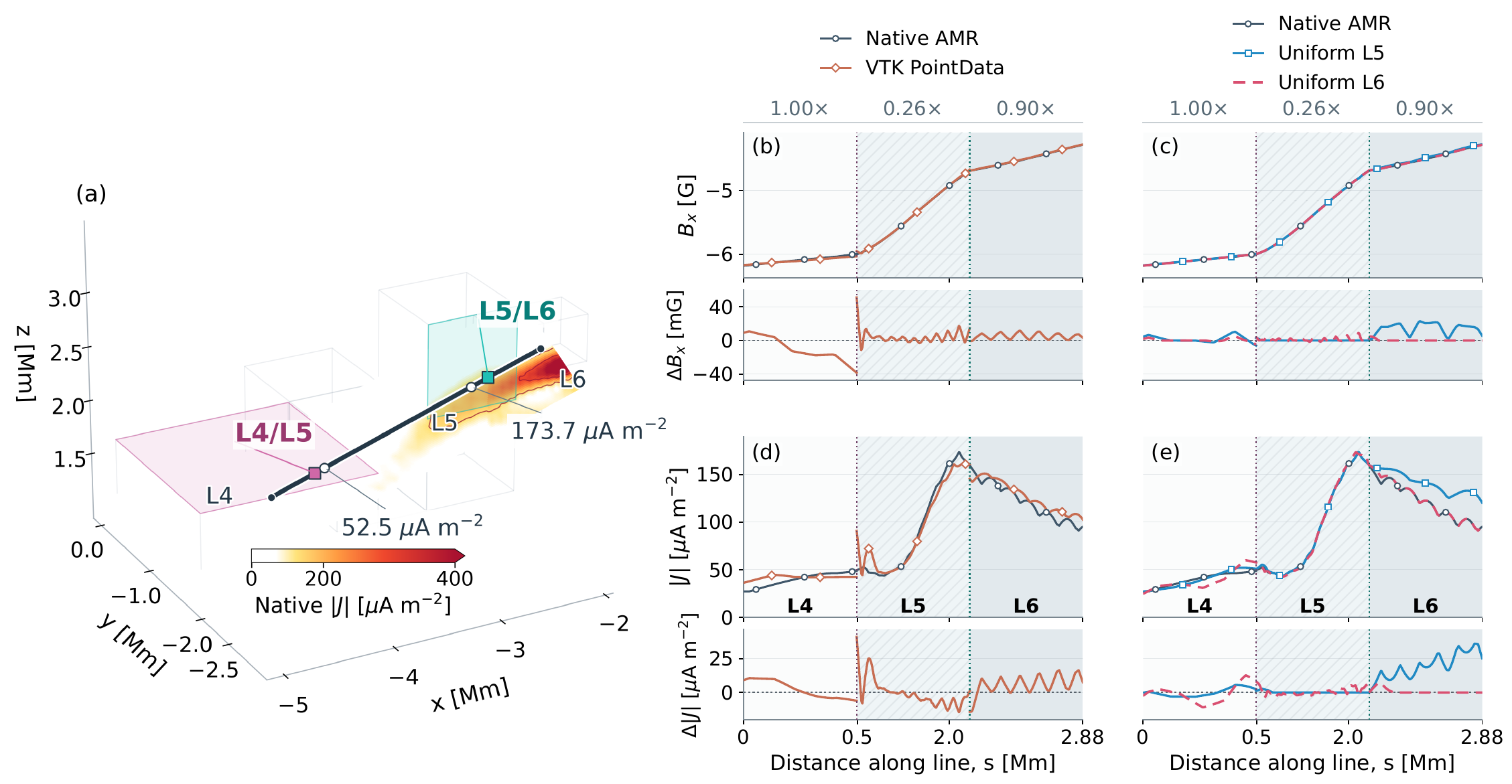}
    \caption{
    Comparison of $B_x$ and $|\mathbf{J}|$ along a sampling line crossing
    native AMR regions at levels L4, L5, and L6.
    (a) Sampling geometry and AMR structure.
    (b, d) Native AMR versus VTK PointData.
    (c, e) Native AMR versus Uniform L5 and Uniform L6.
    Narrow panels show method-minus-native differences.
    The native L4, L5, and L6 intervals occupy equal display widths. The multipliers above the profiles give the horizontal display scale relative to L4, while distance ticks retain their physical values. Visual slopes across these intervals therefore cannot be compared directly as physical gradients.
    }
    \label{fig:field-current-comparison}
\end{figure*}

The comparison is performed along the sampling line shown in Fig.~\ref{fig:field-current-comparison}(a). The line crosses native AMR regions at refinement levels L4, L5, and L6, including the L4/L5 and L5/L6 interfaces. Using the same physical path for all processing routes allows differences in field interpolation and derivative-based diagnostics to be compared under identical sampling conditions. The following analysis therefore examines both the behavior within each refinement region and the changes introduced near coarse--fine interfaces.

\subsubsection{Native AMR and VTK point-data routes}

We first compare the Native AMR and VTK PointData routes. VTK provides the data model and numerical filters used by widely adopted scientific-visualization tools, including ParaView and VisIt, and therefore represents a common post-processing pathway. 

In the Native AMR route, the magnetic field remains cell-centered on each leaf block. Two valid ghost-cell layers are prepared with the \texttt{exact-phase} scheme to supply neighboring values at same-level block boundaries and coarse--fine interfaces. The field at a query point $\mathbf{x}$ is evaluated from neighboring cell-center values by trilinear interpolation,
\begin{equation}
\mathbf{B}_N(\mathbf{x})
=
\sum_{\boldsymbol{\alpha}\in\{0,1\}^3}
w_{b,\boldsymbol{\alpha}}(\mathbf{x})
\mathbf{B}_{b,\boldsymbol{\alpha}},
\label{eq:native-field-query}
\end{equation}
where $b$ denotes the leaf block containing the query point. The interpolation stencil may include valid ghost-cell values.

Numerical magnetic fields may contain nonzero divergence \citep{Chen2026}, and component-wise trilinear interpolation does not enforce $\nabla\cdot\mathbf{B}=0$. For computational efficiency, we use this local interpolation scheme without an additional divergence-free reconstruction or projection.

The current density is evaluated before the field is sampled along the line. The three-dimensional curl is first computed at native cell centers. For magnetic-field component $B_k$ and coordinate direction $d$, the derivative on block $b$ is
\begin{equation}
D_d^{(b)}B_k(\mathbf{i})
=
\frac{
B_k(\mathbf{i}+\mathbf{e}_d)
-
B_k(\mathbf{i}-\mathbf{e}_d)
}{
2h_{b,d}
},
\label{eq:native-centered-difference}
\end{equation}
where $h_{b,d}$ is the local grid spacing of block $b$. The current vector is then formed from the six cross derivatives,
\begin{equation}
\mathbf{J}_{b,\mathbf{i}}
=
\frac{1}{\mu_0}
\begin{pmatrix}
D_y^{(b)}B_z-D_z^{(b)}B_y\\
D_z^{(b)}B_x-D_x^{(b)}B_z\\
D_x^{(b)}B_y-D_y^{(b)}B_x
\end{pmatrix}_{\mathbf{i}}.
\label{eq:native-current}
\end{equation}
Its magnitude is evaluated at the native cell centers and only then sampled along the prescribed line,
\begin{equation}
J_N(\mathbf{x})
=
\sum_{\boldsymbol{\alpha}}
w_{b,\boldsymbol{\alpha}}(\mathbf{x})
\left|
\mathbf{J}_{b,\boldsymbol{\alpha}}
\right|.
\label{eq:native-current-query}
\end{equation}
The Native AMR route therefore preserves the original AMR data locations and evaluates spatial derivatives with the local grid spacing of each block.

The VTK PointData route first changes the data association. The original cell-centered magnetic field is converted to values on shared mesh vertices using \texttt{vtkCellDataToPointData}. At a vertex $p$, the field is constructed from the cells incident on that vertex,
\begin{equation}
\overline{\mathbf{B}}_p
=
\frac{1}{|\mathcal{C}(p)|}
\sum_{c\in\mathcal{C}(p)}
\mathbf{B}_c,
\label{eq:vtk-cell-to-point}
\end{equation}
where $\mathcal{V}(c)$ is the vertex set of cell $c$ and $\mathcal{C}(p)=\{c\mid p\in\mathcal{V}(c)\}$ is the set of cells that include $p$ as a vertex in the converted mesh topology. These are the incident cells used in the average. An interior vertex of a regular hexahedral mesh has eight incident cells; at physical boundaries and coarse--fine interfaces, the set is determined by the actual vertex connectivity. This operation constructs a vertex-based field from the available cell-centered values. The field at an interior point of a hexahedral cell is then interpolated from these vertex values,
\begin{equation}
\mathbf{B}_V(\mathbf{x})
=
\sum_{p\in\mathcal{V}(c(\mathbf{x}))}
N_{c(\mathbf{x}),p}(\mathbf{x})
\overline{\mathbf{B}}_p,
\label{eq:vtk-field-query}
\end{equation}
where $N_{c,p}$ denotes the dimensionless hexahedral shape function associated with vertex $p$ of cell $c$. For the present mesh, this interpolation is locally trilinear. The interpolation acts on the constructed vertex field instead of the prepared native cell-centered field.

The VTK PointData current is obtained by differentiating this PointData representation with \texttt{vtkGradientFilter}. Within a hexahedron $c$, let $B_{V,k}^{(c)}$ denote the $k$th component of the interpolated magnetic field in Eq.~\eqref{eq:vtk-field-query}. The vertex values $\overline{B}_{r,k}$ are fixed coefficients, so spatial differentiation acts on the shape functions. With \texttt{FasterApproximation=False}, the contribution from cell $c$ to the gradient at vertex $p$ is
\begin{equation}
\begin{aligned}
G_{kd}^{(c)}(p)
&=\left.\frac{\partial B_{V,k}^{(c)}}{\partial x_d}\right|_{\mathbf{x}_p}\\
&=\sum_{r\in\mathcal{V}(c)}
\overline{B}_{r,k}
\left.\frac{\partial N_{c,r}}{\partial x_d}\right|_{\mathbf{x}_p}.
\end{aligned}
\label{eq:vtk-cell-gradient}
\end{equation}
Here, $k$ labels the magnetic-field component and $d$ the Cartesian coordinate direction. Although $N_{c,r}$ is dimensionless, its derivative with respect to the physical coordinate $x_d$ has units of inverse length, giving $G_{kd}^{(c)}$ units of magnetic field per unit length. The vertex gradient $\overline{G}_{kd}(p)$ is obtained by averaging these contributions over all incident cells,
\begin{equation}
\overline{G}_{kd}(p)
=
\frac{1}{|\mathcal{C}(p)|}
\sum_{c\in\mathcal{C}(p)}
G_{kd}^{(c)}(p).
\label{eq:vtk-gradient-average}
\end{equation}

The Native AMR and VTK PointData routes differ in both field representation and differentiation. In the Native AMR route, the magnetic field remains cell-centered and spatial derivatives are evaluated on each block using its local grid spacing. In the VTK PointData route, the field is first converted to shared mesh vertices and differentiated on the resulting vertex-based representation. At coarse--fine interfaces, coincident vertices are shared, but no additional hanging-node constraints are imposed; continuity of the PointData field or its gradient is therefore not enforced.

These representation differences are already visible in the sampled magnetic field. As shown in Fig.~\ref{fig:field-current-comparison}(b), the two \(B_x\) profiles remain close over most of the sampling line, with small differences within refinement regions and jumps in the VTK PointData profile at coarse--fine interfaces. These jumps are consistent with the preceding cell-to-point conversion and vertex-based interpolation.

The differences become more pronounced in the current-density magnitude in Fig.~\ref{fig:field-current-comparison}(d). Differentiation increases the sensitivity to local variations in the reconstructed field, making both the interface jumps and the mismatch between the two current profiles more evident. The current-density differences therefore contain contributions from both the field representation and the discrete differentiation used in post-processing, in addition to the discretization already present in the simulation. 

\subsubsection{Uniform-grid resampling and differentiation scale}

The Uniform L5 and Uniform L6 routes provide a complementary comparison. Unlike the VTK PointData route, they retain a cell-centered representation but resample the magnetic field from the native AMR hierarchy onto a uniform target grid. For target level $\ell\in\{5,6\}$, the prepared field is first evaluated at the target-grid centers,
\begin{equation}
\mathbf{B}_{\mathbf{i}}^{(\ell)}
=
\mathbf{B}_N
\left(
\mathbf{x}_{\mathbf{i}}^{(\ell)}
\right).
\label{eq:uniform-resampling}
\end{equation}
This \texttt{simesh}-based point-sampling operation defines the native-to-uniform step used here. The target spacings are $0.10417$ and $0.05208\,\mathrm{Mm}$ for L5 and L6, respectively. Direct evaluation at these centers avoids the grid reconstruction and associated intermediate storage of a full refine/coarsen conversion. A second trilinear interpolation is applied when the resulting uniform field is sampled along the prescribed line,
\begin{equation}
\mathbf{B}_{\ell}(\mathbf{x})
=
\sum_{\boldsymbol{\beta}\in\{0,1\}^3}
v_{\boldsymbol{\beta}}^{(\ell)}(\mathbf{x})
\mathbf{B}_{\boldsymbol{\beta}}^{(\ell)}.
\label{eq:uniform-field-query}
\end{equation}
Uniform L5 and Uniform L6 use the same trilinear procedure at both stages and differ only in target-grid spacing. Inside a same-level block, native-aligned target centers recover the stored cell-center values; equality need not extend to target stencils crossing a refinement interface.

The current density is evaluated after the new magnetic-field representation has been constructed. Spatial derivatives are computed on the uniform target grid,
\begin{equation}
D_d^{(\ell)}B_k^{(\ell)}(\mathbf{i})
=
\frac{
B_k^{(\ell)}(\mathbf{i}+\mathbf{e}_d)
-
B_k^{(\ell)}(\mathbf{i}-\mathbf{e}_d)
}{
2h_{\ell,d}
}.
\label{eq:uniform-centered-difference}
\end{equation}
The full three-dimensional current vector is then formed from these derivatives, and its magnitude is evaluated at the target-grid centers. Only afterward is the scalar current magnitude interpolated to the sampling line,
\begin{equation}
J_\ell(\mathbf{x})
=
\sum_{\boldsymbol{\beta}}
v_{\boldsymbol{\beta}}^{(\ell)}(\mathbf{x})
\left|
\mathbf{J}_{\boldsymbol{\beta}}^{(\ell)}
\right|.
\label{eq:uniform-current-query}
\end{equation}
The Native AMR and uniform-grid processing chains can therefore be summarized as
\begin{equation}
J_N
=
\frac{1}{\mu_0}I_N\left|C_N\mathbf{B}\right|,
\qquad
J_\ell
=
\frac{1}{\mu_0}I_\ell\left|C_\ell R_\ell\mathbf{B}\right|,
\label{eq:operator-chains}
\end{equation}
where $R_\ell$ denotes native-to-uniform resampling, $C$ the discrete three-dimensional curl, and $I$ the interpolation of the already-computed current magnitude. Before the final line query, $C_N$ acts directly on the field using the local block spacing, whereas $C_\ell$ acts on a resampled field with a single target-grid spacing. This difference in differentiation scale is the essential distinction between the two processing routes.

Fig.~\ref{fig:field-current-comparison}(c) shows that both uniform-grid representations reproduce the Native AMR $B_x$ profile closely over much of the sampling line, with small residual differences. These differences arise from the change in interpolation support introduced by resampling. This is particularly relevant near coarse--fine interfaces, where nearby target centers can draw on different native interpolation stencils. Subsequent interpolation from the target grid to the sampling line can therefore differ from a direct native-field query at the same position, even though both routes use trilinear interpolation.

Fig.~\ref{fig:field-current-comparison}(e) shows that these small differences in the sampled magnetic field become more apparent in the current-density profiles. Within the native L4 region, both uniform grids sample more densely than the original mesh, yet Uniform L6 exhibits larger deviations from the Native AMR current than Uniform L5, particularly near the coarse--fine interface. Conversely, within the native L6 region, the coarser Uniform L5 representation yields systematically higher current magnitudes along much of the sampled segment, whereas Uniform L6 closely follows the Native AMR result. The deviations of Uniform L5 in native L6 regions and Uniform L6 in native L4 regions arise from resampling-induced changes in the field differences across the stencil and the spacing used to differentiate them.

These differences reflect changes in both the interpolation support and the spatial scale of differentiation. Resampling changes which native cells contribute to the target values, while the subsequent curl calculation uses the target-grid spacing rather than the local native spacing. Together, these changes account for the sensitivity of the current profiles to target resolution, including the deviations near coarse--fine interfaces.

\section{Applications}
\label{sec:applications}

The following applications use prepared fields on the native AMR representation, with visualizations constructed from the resulting analysis products.

\subsection{Magnetic topology and structure}
\label{sec:magnetic-topology}

We use two simulations to demonstrate magnetic-topology and magnetic-structure applications. The first is the current-sheet simulation of \citet{Wu2025}, also used above to illustrate the native AMR representation. The second is a quiescent solar-filament simulation, which provides a larger-scale solar environment for examining magnetic connectivity and three-dimensional field structure.

\begin{figure*}[!tp]
\centering
\includegraphics[width=\textwidth]{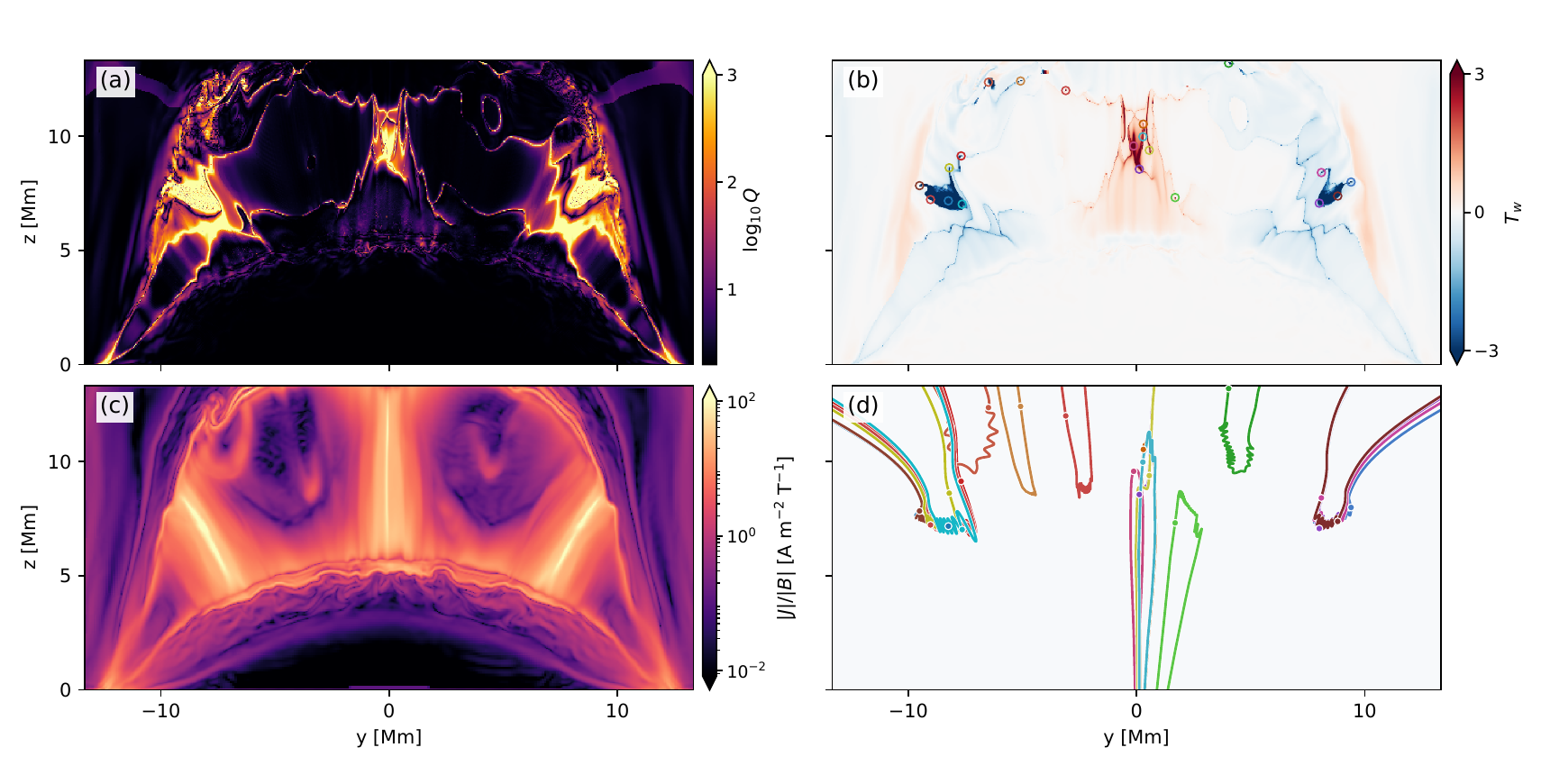}
\caption{Magnetic diagnostics in the current-sheet snapshot.
(a) $\log_{10}Q$ from finite-difference endpoint mapping, (b) signed $T_w$, and (c) $|\mathbf{J}|/|\mathbf{B}|$ on $x=0$. The $Q$ and $T_w$ maps use current-guided adaptive sampling.
(d) Twenty three-dimensional field lines traced from seeds above the 99th percentile of valid $|T_w|$, projected onto the $y$--$z$ plane along $x$. Colored markers in (b) identify the seeds of the correspondingly colored lines in (d).
}
\label{fig:current-sheet-x0}
\end{figure*}

\subsubsection{Quasi-separatrix layers}
\label{sec:qsl-application}

The squashing factor $Q$ measures the distortion of the magnetic field-line mapping between two boundary patches; large values identify quasi-separatrix layers (QSLs), where connectivity varies rapidly. The mapping connects a launch patch $S_1$, with footpoint coordinates $(x_1,x_2)$, to a target patch $S_2$, with conjugate footpoint coordinates $(X_1,X_2)$, along the same magnetic field lines \citep{Titov2007}. These patches may lie on different domain faces or on the same physical boundary, as for the two photospheric footpoints of a closed coronal field line. The field must cross each patch at the corresponding footpoint, with $\mathbf{B}\cdot\mathbf{n}_i\ne0$ for the local surface normal $\mathbf{n}_i$; perpendicular intersection is not required. For seeds inside the volume, tracing in both directions locates the two boundary footpoints.

By default, \texttt{simesh} evaluates $Q$ using variational transport \citep{Scott2017,Chen2026}. Writing $\mathbf{b}=\mathbf{B}/|\mathbf{B}|$ and parameterizing a field line by arc length $s$, two initially transverse perturbations $\mathbf{u}$ and $\mathbf{v}$ obey
\begin{equation}
\frac{d\mathbf{u}}{ds}=(\mathbf{u}\cdot\nabla)\mathbf{b},
\qquad
\frac{d\mathbf{v}}{ds}=(\mathbf{v}\cdot\nabla)\mathbf{b}.
\label{eq:q-transport}
\end{equation}
Their endpoint projections along the magnetic field onto the tangent planes of $S_1$ and $S_2$ determine the mapping Jacobian $D=\partial(X_1,X_2)/\partial(x_1,x_2)$, expressed in local orthonormal tangent coordinates, from which
\begin{equation}
Q=\frac{D_{11}^{2}+D_{12}^{2}+D_{21}^{2}+D_{22}^{2}}{|\det D|}.
\label{eq:q-definition}
\end{equation}
A footpoint finite-difference method is also available \citep{Pariat2012} and is used for the current-sheet example in Fig.~\ref{fig:current-sheet-x0}(a). This $x=0$ section reveals fine connectivity structure within the current sheet, including narrow high-$Q$ layers surrounding localized features. The $Q$ seeds lie on a sampling plane adaptively refined according to the current-density magnitude; the magnetic field and its curl are evaluated on the native AMR representation.

Fig.~\ref{fig:magnetic-topology}(a) shows the filament's connectivity structure using variational transport. In this example, seeds lie at native cell centers on the selected plane, revealing narrow, extended high-$Q$ features in the simulated solar atmosphere.

\subsubsection{Magnetic twist}
\label{sec:twist-application}

The squashing factor $Q$ identifies regions where field-line connectivity changes rapidly. Magnetic twist characterizes twisting along field lines and provides a criterion for selecting tracing seeds within strongly twisted structures. We compute the twist number \citep{Berger2006}
\begin{equation}
T_w=\frac{1}{4\pi}\int_{\mathcal{L}}
\frac{(\nabla\times\mathbf{B})\cdot\mathbf{B}}{|\mathbf{B}|^{2}}\,ds,
\label{eq:twist-definition}
\end{equation}
where $\mathcal{L}$ is the traced field line between its boundary endpoints. The curl is evaluated on the native AMR representation and sampled together with the magnetic field along the integration path. Figs.~\ref{fig:current-sheet-x0}(b) and \ref{fig:magnetic-topology}(b) show the signed twist diagnostic for the current-sheet and filament simulations, respectively, using the same sampling positions as their corresponding QSL maps. We use the magnitude of $T_w$ to select strongly twisted magnetic structures, including candidate flux-rope regions, by identifying locations satisfying
\[
|T_w| > P_{99}\!\left(|T_w|\right),
\]
and use the selected positions as tracing seeds for the subsequent field-line analysis. 

\subsubsection{Magnetic field-line analysis}
\label{sec:field-line-application}

The $|\mathbf{J}|/|\mathbf{B}|$ map in Fig.~\ref{fig:current-sheet-x0}(c) highlights the current sheet but does not clearly reveal its embedded flux-rope structures. Applying the $|T_w|$ criterion from Sect.~\ref{sec:twist-application} provides seeds for tracing the magnetic field through these structures. Fig.~\ref{fig:current-sheet-x0}(d) shows twenty field lines selected in this way, which trace small-scale flux ropes in the current sheet. Each colored marker in panel~(b) is the tracing seed of the same-colored line in panel~(d). These trajectories are integrated in three dimensions and displayed as projections onto the $y$--$z$ plane along $x$.

The same twist-guided selection traces the main filament flux rope in Fig.~\ref{fig:magnetic-topology}(c). From 40 field lines whose seeds were randomly selected with probabilities proportional to $|T_w|$ above its 99th percentile, we retain the 34 lines with apex heights above $40\,\mathrm{Mm}$. These lines outline an extended, twisted magnetic structure, connecting the strong-twist regions in panel~(b) to the three-dimensional geometry of the rope. The lines are colored by the local current proxy $j_*=|\nabla\times\mathbf{B}|$.

\subsubsection{Current-based emission proxy}
\label{sec:emission-proxy-application}

\begin{figure*}[!tp]
\centering
\includegraphics[width=\textwidth,height=0.75\textheight,keepaspectratio]{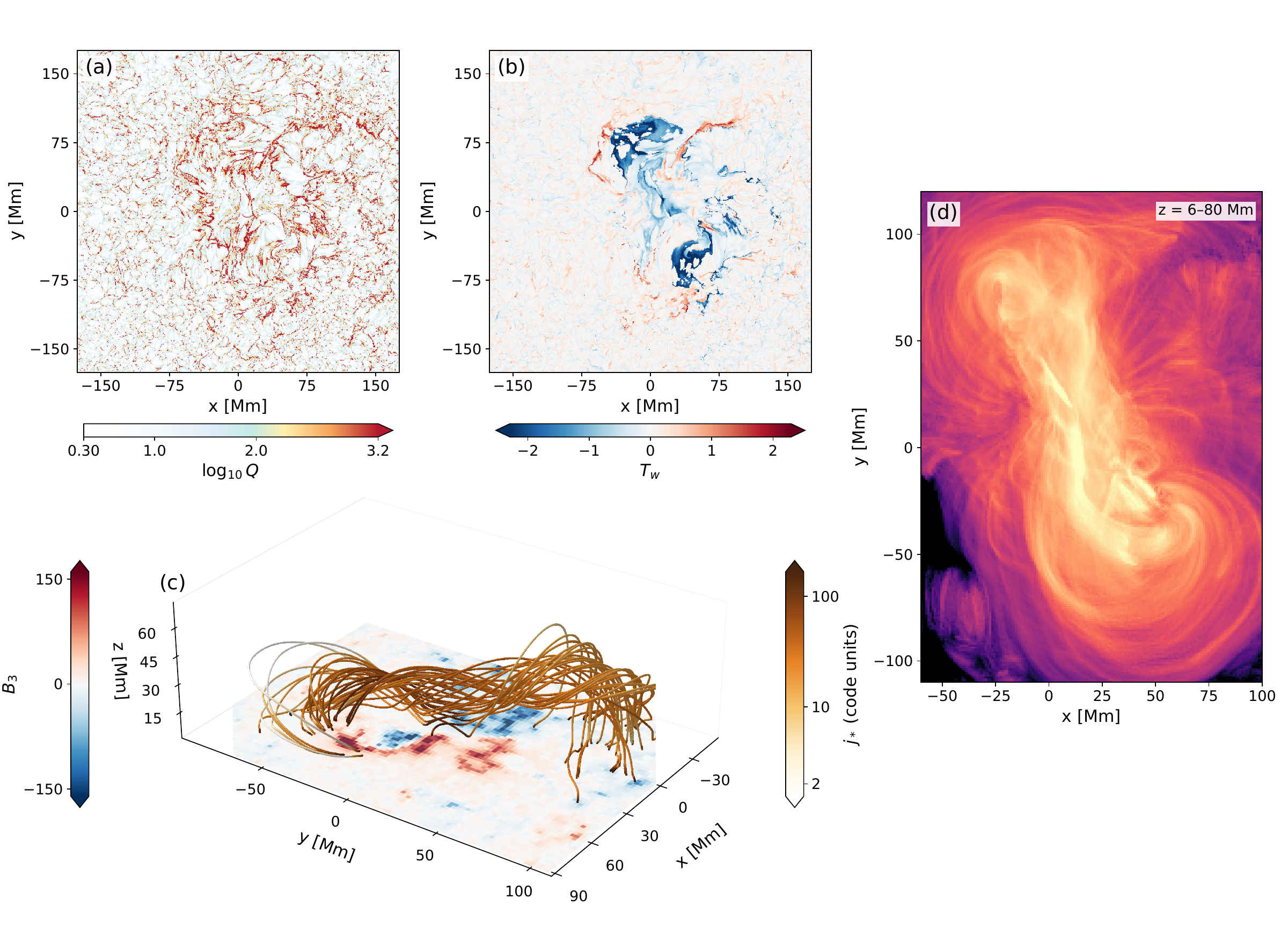}
\caption{Magnetic topology and structure in the filament simulation.
(a) $\log_{10}Q$ from variational transport and (b) signed $T_w$ at $z\simeq2.10\,\mathrm{Mm}$.
(c) Selected field lines colored by $j_*=|\nabla\times\mathbf{B}|$ in code units, above a map of the vertical magnetic-field component $B_3\equiv B_z$.
(d) Current-based emission proxy integrated along $z=6$--$80\,\mathrm{Mm}$.}
\label{fig:magnetic-topology}
\end{figure*}

To visualize the projected magnetic structure, we use the proxy emissivity of \citet{Cheung2012}. This prescription distributes the field-line mean of the squared current over the cells crossed by each closed magnetic field line. With $j_*=|\nabla\times\mathbf{B}|$ evaluated in code units, we compute
\begin{align}
\langle j_*^2\rangle_i &= \frac{1}{L_i}\int_{\mathcal{L}_i} j_*^2\,ds,
\label{eq:mean-current}\\
\epsilon_{*,k} &= G\sum_{i:\,\mathcal{L}_i\cap V_k\ne\varnothing}
A_i\langle j_*^2\rangle_i,
\label{eq:proxy-emissivity}
\end{align}
where $\mathcal{L}_i$ is a bottom-to-bottom magnetic field line of length $L_i$, $V_k$ is a display cell, and $A_i$ is the area represented by the seed on the bottom boundary. We choose the spatially constant coefficient $G=1$. Fig.~\ref{fig:magnetic-topology}(d) shows the line-of-sight integral $I_*(x,y)=\int\epsilon_*(x,y,z)\,dz$. The LOS direction can be chosen arbitrarily, with the $z$ direction used in this example. The proxy provides a morphology diagnostic, without a density, temperature, or instrument-response model for calibrated emission. It is particularly useful for models that omit a full thermodynamic treatment, including zero-$\beta$ models and models with isothermal or polytropic closures.

\subsection{Thermal diagnostics and synthetic observations}
\label{sec:thermal-diagnostics}

\texttt{simesh} provides applications for plasma thermal diagnostics and synthetic observations. We demonstrate these applications using a magnetic eruption simulation.

\subsubsection{Projected quantities and regional statistics}
\label{sec:thermal-statistics-application}

Fig.~\ref{fig:thermal-diagnostics}(a)--(b) shows the column emission measure and emission-measure-weighted temperature along the same line of sight, defined as
\begin{equation}
\begin{aligned}
\mathrm{EM}_{\mathrm{col}}&=\int n_e n_{\mathrm{H}}\,d\ell,\\
\langle T\rangle_{\mathrm{EM}}&=\frac{\int n_e n_{\mathrm{H}}T\,d\ell}{\mathrm{EM}_{\mathrm{col}}}.
\end{aligned}
\label{eq:thermal-projections}
\end{equation}
where $n_e$ and $n_{\mathrm{H}}$ are the electron and hydrogen nuclei number densities under the fully ionized H/He convention, and $\ell$ measures distance along the line of sight through the full $y$ extent of the domain. We adopt $n_{\mathrm{He}}/n_{\mathrm{H}}=0.1$, giving $n_e=1.2n_{\mathrm{H}}$ and $n_{\mathrm{H}}=\rho/(1.4m_p)$, where $m_p$ is the proton mass. The integrands are evaluated at native cell centers before ghost-cell preparation and line-of-sight integration. The column emission measure, in $\mathrm{cm}^{-5}$, shows the spatial distribution of the density weighting, while $\langle T\rangle_{\mathrm{EM}}$ summarizes the temperature along each ray. Their combination distinguishes a high weighted temperature from a large column emission measure.

To compare two height intervals in the central part of the eruption, we select the equal-volume regions R1 and R2 marked in both maps. Both span $x=[-40,40]\,\mathrm{Mm}$ and the full $y$ domain, with $z=[20,60]\,\mathrm{Mm}$ for R1 and $z=[60,100]\,\mathrm{Mm}$ for R2. Fig.~\ref{fig:thermal-diagnostics}(c) resolves their temperature content beyond the projected mean. We construct each distribution from the volume emission measure in logarithmic temperature bins,
\begin{equation}
\Delta\mathrm{EM}_{V,k}=\sum_{i:\,T_i\in[T_k,T_{k+1})}
n_{e,i}n_{\mathrm{H},i}\,\Delta V_{i\cap\mathcal{R}},
\label{eq:thermal-emission-measure-bins}
\end{equation}
where $\Delta V_{i\cap\mathcal{R}}$ is the overlap volume of native cell $i$ with the selected region $\mathcal{R}$. The plotted quantity is $\Delta\mathrm{EM}_{V,k}/\Delta\log_{10}T$, with units of $\mathrm{cm}^{-3}$. This is a volume-integrated emission-measure distribution per logarithmic temperature interval, whereas a line-of-sight differential emission measure (DEM) is defined per unit temperature using a path integral. The summed bin emission measures agree with independent volume integrals.

\subsubsection{Synthetic observations}
\label{sec:synthetic-observations-application}

\begin{figure*}[!tp]
\centering
\includegraphics[width=\textwidth,height=0.84\textheight,keepaspectratio]{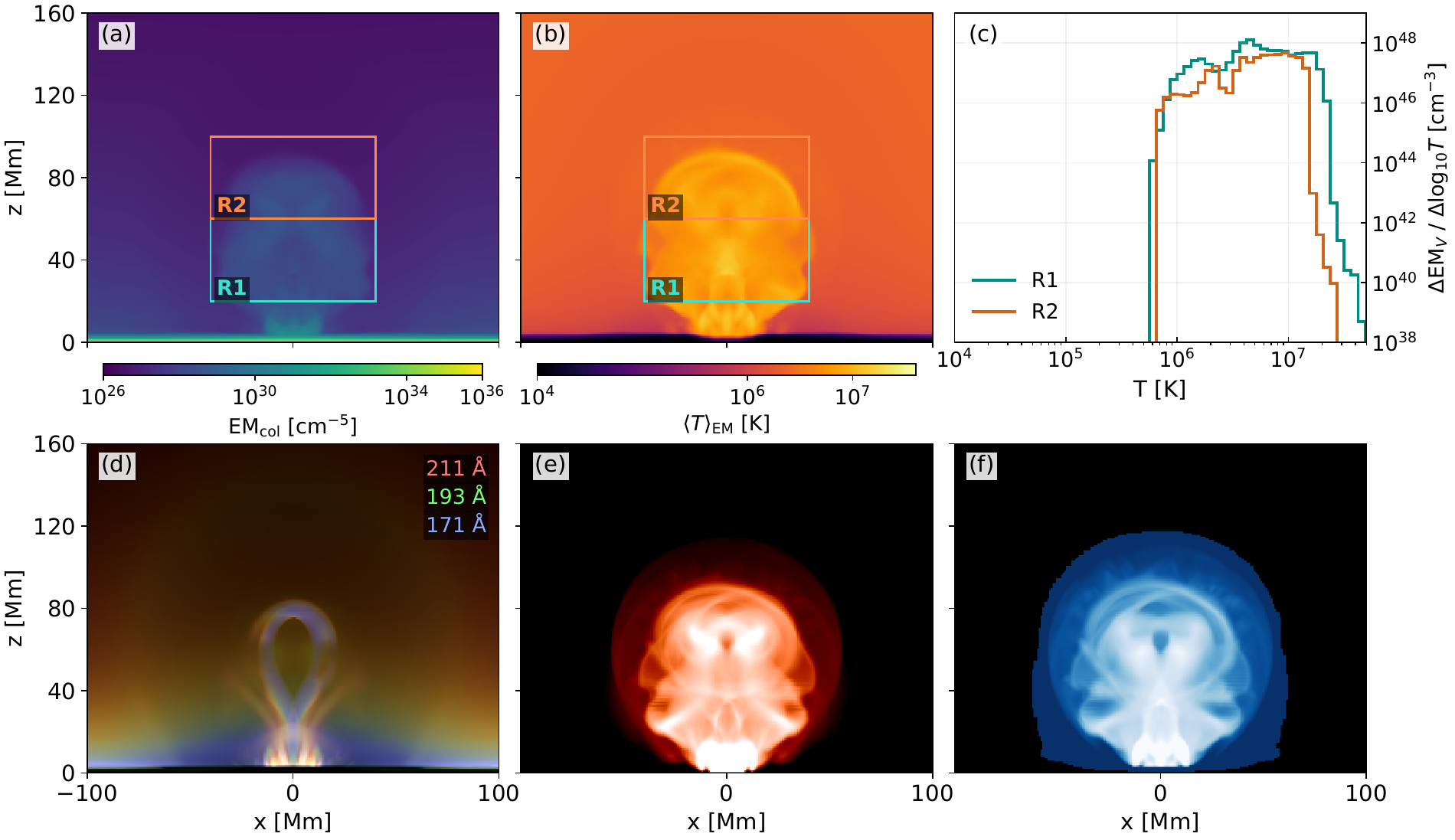}
\caption{Regional thermal diagnostics and synthetic observations from a magnetic eruption simulation.
(a)--(b) Column emission measure and emission-measure-weighted temperature integrated through $y=-200$ to $200\,\mathrm{Mm}$.
(c) Volume emission measure per logarithmic temperature interval for R1 and R2.
(d) AIA 211/193/171~\AA\ RGB composite using independently scaled channels.
(e) IRIS 1354~\AA\ intensity with a warm color table.
(f) EIS 192~\AA\ intensity with a blue color table.
}
\label{fig:thermal-diagnostics}
\end{figure*}

Fig.~\ref{fig:thermal-diagnostics}(d)--(f) shows synthetic emission maps viewed from $y=-200\,\mathrm{Mm}$ along $+y$. For a selected passband or spectral line, the emergent intensity is computed as
\begin{equation}
I_{\lambda}=\int_0^L j_{\lambda}(\ell)\exp[-\tau_{\lambda}(\ell)]\,d\ell,
\label{eq:synthetic-intensity}
\end{equation}
where $\ell$ increases from the observer-side boundary through the domain of length $L$, $j_{\lambda}=n_e n_{\mathrm{H}}R_{\lambda}(T)$, and $\tau_{\lambda}(\ell)=\int_0^{\ell}\kappa_{\lambda}(s)\,ds$ is the foreground optical depth. The emitting densities use the same fully ionized H/He convention as Eq.~\ref{eq:thermal-projections}; $R_{\lambda}(T)$ is the tabulated temperature response, and $\kappa_{\lambda}$ is the absorption coefficient. We adopt the response tables provided by the MPI-AMRVAC thermal-emission module for AIA \citep{Lemen2012}, IRIS \citep{DePontieu2014}, and EIS \citep{Culhane2007}. Emissivity and opacity are evaluated at native cell centers before ghost-cell preparation and spatial interpolation. Ordered integration composes constant-coefficient slab solutions along each ray. Setting $\kappa_{\lambda}=0$ recovers the optically thin intensity.

The AIA composite in Fig.~\ref{fig:thermal-diagnostics}(d) maps 211, 193, and 171~\AA\ to red, green, and blue, respectively. Each channel is logarithmically scaled between fixed limits, clipped to $[0,1]$, and squared to reduce the display contribution of diffuse background emission. The independently normalized RGB channels allow comparison of the spatial morphology in the three passbands. Fig.~\ref{fig:thermal-diagnostics}(e)--(f) shows separate IRIS 1354~\AA\ and EIS 192~\AA\ intensity maps with warm and blue color tables, respectively, and independent logarithmic display ranges. Both maps show extended emission in the central region. These response-weighted projections complement the column emission measure and weighted-temperature maps. The workflow also supports AIA 94, 131, 304, and 335~\AA\ and EIS 255, 263, and 264~\AA\ through their respective response tables.

The EUV maps include photoionization absorption by H~I, He~I, and He~II, which attenuates radiation from emitting plasma behind cool, dense material. Absorbing populations are estimated using LTE Saha equilibrium at the local density and temperature, while the emitting density convention remains fully ionized H/He. The three continua contribute below 912, 504, and 228~\AA, respectively; IRIS 1354~\AA\ therefore uses the optically thin limit. Optical depth and unabsorbed intensity are retained to quantify the effect of absorption.

\section{Discussion and conclusions}
\label{sec:discussion-conclusions}
We have developed \texttt{simesh}, an independent Python framework for quantitative analysis of native AMR data. It retains the native AMR hierarchy, local cell scales, and cell-centered field organization while explicitly preparing the ghost-cell values required by each numerical scheme. Interpolation, spatial derivatives, and tracing then operate directly on the representation. 

The framework also places the four groups of scientific applications on the same numerical basis. In \texttt{simesh}, these operations use the same simulation fields, native AMR hierarchy, and analysis geometry. Their outputs can also be reused directly in later steps. A diagnostic map may define a region of interest, a selected geometry may provide sampling locations, and a retained trajectory may be used to extract additional plasma quantities. This organization reduces the need to move repeatedly between different data models and makes multi-stage analysis easier to construct and reproduce.

The numerical comparisons quantify the effects of representation choice. The Native AMR, VTK PointData, Uniform L5, and Uniform L6 routes do not apply identical data associations, spatial scales, or derivative stencils. These differences may remain small in a sampled primary field but become much larger in derivative-based quantities. The effect is particularly important near coarse--fine interfaces, where neighboring values and grid spacing change. For analyses that retain the simulation's discrete representation, the Native AMR route avoids additional data-association and uniform-resampling steps. At the same time, the framework provides a controlled environment for testing alternative ghost-cell preparation schemes, interpolation rules, and differential operators. Such comparisons can help identify numerical sensitivities and provide feedback for both post-processing methods and simulation algorithms.

MPI-AMRVAC offers another route to uniform-grid data by refining or coarsening a snapshot to a common level through \texttt{level\_io}, then extracting the resulting cell-centered values. This avoids the additional interpolation used to sample the uniform target grids in our comparison. However, the required refinement, coarsening, and storage can be substantial for large three-dimensional snapshots, particularly at high target levels.

MPI-AMRVAC already supports many of the analyses presented here, including field-line tracing and thermal-emission synthesis. Its MPI block decomposition is designed for simulation updates, whereas streamlines can traverse blocks assigned to several processes, requiring communication and synchronization as tracing proceeds. In \texttt{simesh}, independent trajectories are distributed among workers that access the prepared fields in shared memory. Custom analyses within MPI-AMRVAC may also require runtime configuration, user routines, or recompilation. For thermal synthesis, the contribution of \texttt{simesh} lies primarily in integrating the existing emission prescriptions with regional statistics, magnetic diagnostics, and visualization in a reusable Python workflow.

\texttt{simesh} currently supports three-dimensional Cartesian AMR data from MPI-AMRVAC and compatible AGILE simulations. Planned extensions include support for spherical and other geometric configurations, as well as data from other AMR frameworks, such as AMR--CESE--MHD \citep{Jiang2010CESE}, and the six-component overset grids used by SIP--CESE MHD and its AMR implementations \citep{Feng2010SIP,Feng2014AMR}. The reusable Python interfaces could also be exposed as callable tools for AI coding agents, allowing them to query simulation fields and request quantitative diagnostics. Their numerical outputs and visualizations would remain available for inspection and subsequent analysis.

\section*{Code availability}
\label{sec:code-data-availability}
The source code of \texttt{simesh} is available under the GNU General Public License v3.0 at \url{https://github.com/Astery0502/simesh}. 

\section*{Acknowledgements}
This work is supported by the Specialized Research Fund for State Key Laboratory of Solar Activity and Space Weather, NSFC 1250030413.
Y.G. was supported by the National Key R\&D Program of China (2022YFF0503004), NSFC (12333009), and the Fundamental Research Funds for the Central Universities (KG202506).

\clearpage
\bibliographystyle{aasjournalv7}
\bibliography{references}

\end{document}